\documentclass[aps,twocolumn,prd]{revtex4}

\usepackage{graphicx}
\DeclareGraphicsExtensions{.pdf}

\newcommand{\beq}{\begin{equation}}
\newcommand{\eeq}{\end{equation}}
\newcommand{\bea}{\begin{eqnarray}}
\newcommand{\eea}{\end{eqnarray}}

\newcommand{\bra}[1]{\left< #1 \right|}
\newcommand{\ket}[1]{\left| #1 \right>}

\newcommand{\dg}{^\dagger}

\newcommand{\Hc}{\mathrm{H.c.}}

\newcommand{\half}{\frac{1}{2}}

\newcommand{\pr}{^\prime}

\newcommand{\om}{\omega}

\newcommand{\vphi}{\varphi}

\newcommand{\pip}{\pi^{+}}
\newcommand{\pin}{\pi^{-}}
\newcommand{\xv}{\mathbf{x}}

\newcommand{\vv}{\mathbf{v}}
\newcommand{\kv}{\mathbf{k}}
\newcommand{\ak}{a_\kv}
\newcommand{\akd}{a_\kv\dg}
\newcommand{\omkv}{\om_\kv}

\newcommand{\Dc}{\mathcal{D}}

\newcommand{\Kb}{\mathbf{K}}
\newcommand{\Kc}{\mathcal{K}}
\newcommand{\Kcb}{\Kc\!\!\!\!\Kc}

\newcommand{\Pb}{\mathbf{P}}
\newcommand{\Pc}{\mathcal{P}}
\newcommand{\Pcb}{\Pc\!\!\!\!\!\Pc}

\begin{document}

\title{Relativistic GKLS master equation?}
\author{Lajos Di\'osi}
\affiliation{Wigner Research Centre for Physics, 
                      H-1525 Budapest 114, P. O. Box 49, Hungary\\
                      and\\
                      E\"otv\"os Lor\'and University, H-1117 Budapest, 
                      P\'azm\'any P\'eter stny. 1/A}
                                                       
\date{\today}
\begin{abstract}
The celebrated GKLS master equation, widely called just Lindblad equation,  
is the universal dynamical equation of non-relativistic open 
quantum systems in their Markovian approximation. It is not necessary
and perhaps impossible that GKLS equations possess sensible relativistic 
forms. In a lucid talk on black hole information loss paradox, David Poulin conjectured a Lorentz invariant GKLS master equation. It remained unpublished. 
Poulin passed away at heights of his activity. But the equation is really puzzling.  
A closer look uncovers a smartly hidden defect which leaves us without Lorentz invariant Markovian master equations. They, in view of the present author,
should not exist. 
\end{abstract}


\maketitle

\section{Introduction}
Dissipative relativistic phenomena are real.  
A natural example are pions. 
If we regard the dynamics of pions in itself, it is relativistic and dissipative:
the pionic state decays toward the pionic vacuum state. 
The dynamics is the reduced dynamics of a unitary quantum field theory  (QFT, Standard Model),
and as such, it is non-Markovian: the time-derivative $d\rho(t)/dt$ 
of the pionic state depends on the history of $\rho$ before $t$. 

Long ago and far from the context of QFT, a very powerful mathematical
theorem \cite{Lin76,GKS76} proved (see also \cite{Fra76,GKLS})
that non-relativistic Markovian evolution of quantum
states can always be expressed by a very specific structure of a number 
of operators $A_n$:
\beq\label{GKLS}
\frac{d\rho}{dt}=-i[H,\rho]+\sum_n \left( A_n\rho A_n\dg-\half\{A_n\dg A_n,\rho\}\right).
\eeq
Popularity of this  GKLS master equation, many times referred just
as Lindblad master equation after one of the inventors, has been
and is remarkably extending in many fields in non-relativistic quantum
physics. It is understood as a Markovian effective equation of
open quantum systems \cite{BrePet02} whose exact dynamics 
is non-Markovian. I used to share this view. The only exact Markovian evolutions
are unitary. Exact non-unitary (e.g. dissipative) Markovian evolutions
do not exist. Lorentz invariant Markovian dissipation is impossible. 

An unexpected push came from David Poulin proposing a relativistic
GKLS equation in his 2017 talk \cite{Pou17}. The proposal is impressive and has been
shaking my firm judgement that relativistic GKLS equations are non-existing.    

\section{Poulin's observation }
Consider a quantized free scalar field $\vphi$ of mass $m$ and its canonical momentum $\pi$.
The Hamiltonian $H$ reads
\bea
H&=&\half\int\left(\pi^2+(\nabla\vphi)^2+m^2\vphi^2\right) d\xv\nonumber\\
   &=&\int\omkv\akd\ak d\kv.\label{H} 
\eea
The state $\rho$ evolves by the
Schr\"odinger (--von-Neumann) equation of motion
\beq 
\frac{d\rho}{dt}=-i[H,\rho].\label{Sch}
\eeq
Lorentz invariance relies simply
on the fact that $H=P_0$ where 
\beq\label{Pmu}
    P_\mu=\int k_\mu\akd\ak d\kv
\eeq
is a 4-vector (of total energy-momentum). 

One can modify the free unitary dynamics by a non-unitary 
(e.g.: dissipative) mechanism represented by a superoperator $\Dc$:
\beq\label{ME}
\frac{d\rho}{dt}=-i[H,\rho]+\Dc\rho,
\eeq
where the dissipator $\Dc$ has the GKLS structure (\ref{GKLS}).
Poulin's proposal is this:
\bea\label{D}
\Dc\rho&=&\gamma\int\left(2\pin\rho\pip-\{\pip\pin,\rho\}\right)d\xv\nonumber\\ 
               &=&\gamma\int\omkv\left(\akd\rho\ak 
                                                        -\half\{\akd\ak,\rho\}\right)d\kv,
\eea
where $\pi^\pm$ are the positive and negative frequency parts of $\pi$.
The argument of Lorentz invariance is the same as above.
One can write $\Dc$ in the form
\beq
\Dc=\gamma\int\omkv\left(\akd\otimes\ak 
                                                        -\half(\akd\ak\otimes I+I\otimes\akd\ak)\right)d\kv
\eeq
and argue that $\Dc=\mathcal{P}_0$ where 
\beq\label{Psupmu}
\mathcal{P}_\mu=g\int k_\mu\left(\akd\otimes\ak 
                                                                 -\half(\akd\ak\otimes I+ I\otimes\akd\ak)\right)d\kv   
\eeq
is a 4-vector. 

With the new dissipative mechanism the bosons are decaying and 
for long time the system's state  becomes the vacuum.
The stable equilibrium vacuum state  is supposed to be approached along
a relativistic invariant Markovian evolution by construction. 
Poulin notes that the dynamics, unlike in standard QFT, is
non-local  on range $1/m$. The resulting acausality is of short range
provided $m$ is large. This can, in certain theories, be a bearable anomaly.

However, the forthcoming analysis  uncovers that  the eq. (\ref{ME})
is not Lorentz invariant. 
The next section formulates the condition of boost invariance in
Markovian dissipative quantum fields, like the proposed one.
A lapse of Poulin's argument is detected. 

\section{Condition of boost invariance}
Let us recapitulate the condition of invariance under Lorentz boosts
in standard QFT, with interaction $V$.  
Let us evolve the system dynamically for a small time $\delta t$ 
and perform a boost 
with small velocity $\delta\vv$. Or, apply the boost first and let 
the system evolve after it. If the dynamics is Lorentz invariant then the resulting 
two states must coincide apart from the spatial shift $\delta\vv\delta t$
in the second  state.  The mathematical condition of
this invariance (i.e.: interchangeability of dynamical evolution and boost)
is the following:  
\beq\label{inv}
[\Kb,H+V]=i\Pb,
\eeq
where $\Kb$ is the generator of boosts and $\Pb$ is the spatial part of
$P_\mu$ in (\ref{Pmu}). The closed expression of $\Kb$ exists \cite{Wei95} 
but in practice we use the boost action on the operator basis $\ak,\ak\dg$. 
The small boost acts like this:
\beq\label{smallboost}
\ak+i\delta\vv[\Kb,\sqrt{\omkv}\ak]=\sqrt{\om_{\kv\pr}}a_{\kv\pr}
\eeq
and similarly for $\ak\dg$, where $\kv\pr=\kv-\delta\vv\omkv$ is
the boosted $\kv$. Hence, the boost of any operator 
is equivalent with the boost of the  (covariant) creation/annihilation operators.
We have $[\Kb,H]=i\Pb$, and $[\Kb,V]=0$ for non-derivative interaction,
the condition (\ref{inv}) is satisfied. 

In the proposed  eq. (\ref{ME}) the Hamiltonian interaction term $-i[V,\rho]$ is 
replaced by the dissipative term $\Dc\rho$.  The second term $[\Kb,V]$ of the
condition (\ref{inv}) becomes non-vanishing:
\beq
(\Kb\otimes I)\Dc-\Dc(I\otimes\Kb)=i\Pcb,
\eeq
where $\Pcb$ is the spatial part of $\Pc_\mu$ in (\ref{Psupmu}).
The condition (\ref{inv}) of boost invariance becomes violated.

\begin{figure}[h]
\includegraphics[width=0.5\textwidth]{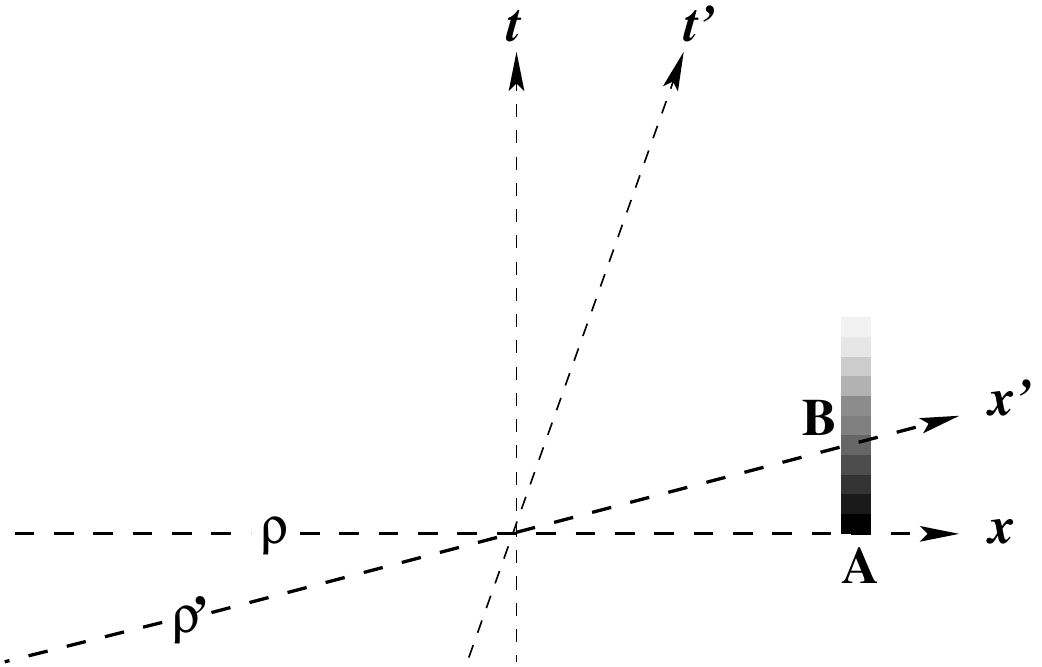}
\caption{In frame $(t,x)$, a single-boson non-relativistic localized state 
is prepared at location A $(t=0,x>0)$ at rest. For $t>0$, the boson is starting
to decay. The initial local system at A reaches B in an irreversible process.
If the initial state $\rho$ defined at $t=0$ were unitary equivalent with
$\rho'$ defined at $t'=0$ ---where $(t',x')$ is a different Lorentz frame---
then the evolution of our local boson should be reversible, which is not the case.}
\label{fig}
\end{figure}

Now we put the argument of Sec. II under scrutiny. 
The proposal assumes that the boost generator is the standard
Hermitian generator $\Kb$, acting as in eq. (\ref{smallboost}). 
This cannot be true.  Since the time-evolution is not unitary the
boosts cannot be unitary either (Fig. \ref{fig})!

The boost generator might become a superoperator $\Kcb$ to satisfy
the condition of invariance,  i.e.: the interchangeability between 
dynamical evolution and boost. The superoperator counterpart of the 
mathematical condition (\ref{inv})  of boost invariance is straightforward.
But, in the next section we show that it is useless to search for 
the covariant boost. The eq. (\ref{ME}) cannot be Lorentz invariant.

\section{Disproof of Lorentz invariance}
The dissipative term does not prevent us from using an interaction picture.
We use an unconventional interaction picture where $H$
evolves the state and $\Dc\dg$ evolves the field:
\bea
\frac{d\rho}{dt}&=&-i[H,\rho],\\
\partial_t \vphi(t,\xv)&=&\Dc\dg \vphi(t,\xv).
\eea
The generator $H$ of the unitary evolution and the generator $\Dc$ of the dissipative evolution are commuting hence  the  constant $H$ governs the
state evolution.  Now, the evolution of the state is standard 
Lorentz invariant. What about the evolution of the field? 
The initial condition reads:
\beq\label{phiIn}
\varphi(0,\xv)=\frac{1}{(2\pi)^{3/2}}
\int\frac{1}{\sqrt{2\omkv}}a_\kv e^{i\kv\xv} d\kv+\Hc~.
\eeq
From the relationships $\Dc\dg a=-\gamma a$ and $\Dc\dg a\dg=-\gamma a\dg$,
the solution follows easily:
\beq\label{phi}
\varphi(t,\xv)=\frac{1}{(2\pi)^{3/2}}
\int\frac{1}{\sqrt{2\omkv}}a_\kv e^{i\kv\xv-\gamma\omkv t}d\kv+\Hc~.
\eeq
One would prove or disprove the boost invariance of the solutions. 
But we have a simpler tool, the field equation:
\beq
\partial_t^2 \vphi(t,\xv)=\gamma^2(m^2-\nabla^2)\vphi(t,\xv),
\eeq
which is manifest non-invariant. This is not surprising since
Sec. III found already a flaw in the argument supporting Lorentz invariance
of the proposal in Sec. II.

\section{Digression: classical and quantum white noise}
A naive Lorentz invariant field theory appeared in \cite{BarLanPro86} first,
where
\beq\label{BLP}
\Dc\rho=g^2\int \left(\vphi\rho\vphi-\half\{\vphi^2,\rho\}\right)d\xv.
\eeq 
This is a Lindblad form (\ref{GKLS}) and the corresponding dynamics
is Lorentz invariant indeed. 
It can be derived from the coupling $g\phi\xi$
to an external Lorentz invariant classical white noise field of ultra-local correlation
\beq
\langle \xi(x)\xi(y)\rangle=g\delta(x-y),
\eeq
after taking the average over this random field. The features of $\Dc$
are unphysical, it is creating bosons at infinite rate which is a trivial
consequence of the white noise. Unfortunately, $\xi(x)$ is the only
possible Lorentz invariant white noise, or, in other words, the only
Lorentz invariant classical Markovian process on the continuum.

We can construct a Lorentz invariant quantum white noise $b(x)$
as well. It is a trivial relativistic generalization of quantum white noise $b(t)$ 
introduced for damped quantum systems \cite{GarCol85} and extensively used 
e.g. in quantum optics \cite{WalMil94}. 
The canonical commutator is ultra-local bosonic:
\beq
[b(x),b\dg(y)]=\delta(x-y).
\eeq  
We use $b(x)$ as an auxiliary field to construct a unitary QFT. 
Poulin's impressive proposal corresponds to the coupling 
\beq\label{coupling}
\sqrt{2\gamma}(\pip b+ \pin b\dg).
\eeq
Assuming that the initial state of the $b$-field is the vacuum state, we evolve the composite state $\rho\otimes\ket{vac}\bra{vac}$ unitarily and
trace out the auxiliary field.  We mentioned in Sec. I that 
in standard QFTs the reduced dynamics are non-Markovian. But
the auxiliary $b$-field is exceptional, it is ultra-local, non-propagating, e.t.c.,
so we get a Markovian evolution for $\rho$ of the $\vphi$-field. 
This is exactly Poulin's dissipative dynamics (\ref{ME}) in interaction picture:
\beq
\frac{d\rho}{dt}=
\gamma\int\left(2\pin\rho\pip-\{\pip\pin,\rho\}\right) d\xv,
\eeq
which is not Lorentz invariant according to Secs. III-IV. 

How is it possible?
The coupling was Lorentz invariant, the reduction is Lorentz invariant, 
then where has Lorentz invariance been lost? Sure,  Lorentz invariance of
the reduced dynamics is undermined
by the non-locality of $\pi^\pm$ in the otherwise Lorentz invariant coupling (\ref{coupling}). Weinberg \cite{Wei95} warns us about the importance of
locality condition. 
\emph{It is this condition that makes the combination of Lorentz
invariance and quantum mechanics so restrictive.}

\section{Closing remarks}
For long time there have been one only context with the interest and
unfulfilled desire for relativistic GKLS equations. The assumption of a 
tiny fundamental and spontaneous decoherence in massive
degrees of freedom was realized by the non-relativistic
GKLS equations \cite{Dio87,GhiPeaRim90}, but the relativistic extensions
are missing up till now. Efforts
 \cite{Dio90,Pea90,KurFra11,Pea99,BedPea19,CaiGasBas21}, 
mostly related to the structure (\ref{BLP}),
are always leading to unphysical features, like, 
e.g., the mentioned vacuum instability, or just presence of tachyons.

Poulin's  motivation was not different in that he assumed a tiny fundamental
dissipative mechanism. He did it directly in the relativistic realm. 
The proposal is smartly hiding its defect. To point it out  took quite 
a time for the present author.

Free pions decay exponantially, they follow a Markovian Lorentz invariant effective dynamics.
Their exact dynamics cannot be Markovian. Any Markovian irreversible field 
process ---whether quantized or classical--- is underlain by  instantaneous 
jumps and they do not exist relativistically.   

\emph{Acknowledgements.} I thank Peter Vecserny\'es for useful discussions.
I acknowledge support from the
Foundational Questions Institute and Fetzer Franklin
Fund, a donor advised fund of Silicon Valley Community
Foundation (Grant No. FQXi-RFP-CPW-2008), the
National Research, Development and Innovation Office
for ``Frontline'' Research Excellence Program (Grant No.
KKP133827) and research grant (Grant. No. K12435),
the John Templeton Foundation (Grant 62099).


\begin{thebibliography}{99}
\bibitem{GKLS} Proofs \cite{Lin76,GKS76} were based on the important
                  restrictive condition of complete-positivity. In \cite{Fra76} this 
                  condition is implicite and its role remains somehow opaque.
\end{thebibliography}
\end{document}